# A Martingale Approach to Detect Peak of News in Social Network

Saba Babakhani[1], Niloofar Mozaffari[2] and Ali Hamzeh[3]

[1] CSE and IT Department, Shiraz University
Shiraz, Fars, Iran

[2] CSE and IT Department, Shiraz University
Shiraz, Fars, Iran

[3] CSE and IT Department, Shiraz University
Shiraz, Fars, Iran

**Abstract**
Nowadays, social medias such as Twitter, Memetracker and Blogs have become powerful tools to propagate information. They facilitate quick dissemination sequence of information such as news article, blog posts, user's interests and thoughts through large- scale. Providing strong means to analyzing social networks structure and how information diffuse through them is essential. Many recent studies emphasize on modeling information diffusion and their patterns to gain some useful knowledge.
In this paper, we propose a statistical approach to online detect peak points of news when spread over social networks, to the best of our knowledge has never investigated before. The proposed model use martingale approach to predict peak points (when news reached the peak of its popularity). Experimental results on real datasets show good performance of our approach to online detect these peak points.

***Keywords:*** *Social Network, Information Diffusion, Martingale, Peak Points.*

## 1. Introduction

Social Network Analysis (SNA) is one of the most important topics for many researchers from different fields. Because of very complete covering on several aspect of human sociality's life, social network analysis and its researches areas are the center of attentions.
Social network like Twitter[1] has revolutionized the way of user interactions with each other and the topics of information, such that social media's users play a significant role to propagate information. Information Diffusion is one of the most important topics in SNA. This research area investigates spreading of information across the network, such as latest news headlines, political movement, movies recommendation, Marketing management, etc.
Gaining a deep understand on how information propagate through the network is an active research area recently [19, 20, 7, 11]. As we know a social network is a graph of nodes with edges between them. Social network structure is dynamic and due to the complexity of this structure, finding patterns of information diffusion is a challenging task. In social network we can see when a node becomes informed (received a piece of information), but we cannot understand where is the source of those information.
There are various methods to finding patterns of information diffusion. The first one is epidemiology theory [24, 26]. Epidemiology theory has an infection probability that use to show the popularity of a news or strength of infection that each news creates. Susceptible – Infected (SI) Model proposed by May and Anderson [26]. Each node in SI model has one of two states (susceptible or infected) and each infected node tries to infect susceptible nodes based on its strength of infection. This model follows rise and fall exponential distribution. YouTube views per a day by users are modeled by Crane and Sornette (C-S) [25]. C-S model has used power law distribution for rise and fall patterns. An implicit model for modeling information diffusion in social networks proposed by Yang et.al [13]. This model uses for implicit networks, where we don't have any explicit knowledge of the network. SpikeM model is the other model that proposed by Mutsabara et.al [28] that uses exponential rise and power law fall.
This paper proposes a statistical approach to online detect peak of news in social networks, to the best of our knowledge has never been applied before. In order to online detect peak of news our approach uses exchangeable test. This hypothesis test is driven from martingale which is based on Doob's Maximal Inequality.

---

[1] www.twitter.com

The rest of this paper goes as follows; the next section gives some background on existing methods for finding and modeling patterns of information diffusion in social networks. Section 3 present the proposed approach and related procedures are defined. The validation of this approach is illustrated in section 4 while the conclusion of this paper provides in section 5.

## 2. Related Work

Diffusion process as a fundamental approach over social networks has attracted many attentions nowadays [6, 21, 18, 29 13]. Modeling information diffusion in social networks and blogs has various viewpoints that we provide a survey on all of them in below parts.
The first and common theory for modeling information diffusion is susceptible- infected [26] where we have nodes in one of two states, susceptible or infected. Infected nodes in the network try to infect their susceptible neighbor nodes by the probability or strength of infection. We have an infection when susceptible nodes faced with an infected node. Patterns of information diffusion by this model follow exponential distribution for rise and fall around the peak points.
Based on the Babarasi research [3, 4, 2], all the patterns are associated with human interactions and blog posting are based on the power law distribution [12, 4, 22]; where the popularity of a post decays over the time. The form of the power law distribution is based on the Eq. (1), where $\alpha = 1.5$, that shows fall patterns of information diffusion in blog post have power law distribution [12].

$$p(n) = n^{-\alpha} \tag{1}$$

A model in 2007 proposed by Leskovek and Backstorm [12] to find out how information disseminate in blog domain. Power law fall patterns of this model shows blog posts lose their popularity over the time.
Crane and Sornette (C-S) [25] used self-excited Hawkes condition Poisson process [1] to model YouTube views in each day. C-S model follows power law rise and fall patterns.
Yang and Leskovec [13] introduced a model for modeling information diffusion in implicit social networks. In implicit network we don't have any explicit knowledge about the network and its structure. Linear influence model LIM) predicts number of infected nodes in future based on the other nodes that infected in the past.
A non-parametric model entitled K-Spectral Centroid (K-SC) presented by Yang and Leskovek in 2011 [14]. Their model introduced for clustering the time series of news spreading and most popular phrases of blog posts in social networks. Most popular phrases extracted based on the Memetracker methodology that introduced by Leskovek et.al [11]. A non-parametric model is not appropriate for modeling information diffusion.
In 2012 SpikeM model proposed by Matsubara et.al [28]. A parametric model that follows exponential rise and power law fall. This model used Levenberg- Marquardt optimization [10] to estimate parameter' value. SpikeM model is better than the other presented above model to modeling patterns of information diffusion.
Time series analysis traditional method in data mining [9] like Auto Regression (AR) [16], Linear Dynamical System (LDS), Kalman Filters (KF) and their variants [2, 17, 15] are not suitable for modeling information diffusion in social networks because of their linear structure.
Some non- linear method are proposed but because of using nearest neighbor rule or artificial neural network [5], they have some obstacle to predict number of infected nodes in the future [5].

## 3. Proposed Method

In order to detect peak of news in social networks, we use our approach that was proposed in [23]. We assume $\{I_1, I_2, I_{n-1}\}$ is a sequence of number of infected nodes in each time step and new informed node $I_n$ is obtained. Our purpose is to discover the peak of news. When each instance (the number of infected nodes in each time) receives, a hypothesis testing happens to figure out whether peak is happened or not. This test is based on exchangeability condition and defined as follow:

$H_0 = \{There\ is\ no\ peak\ points\ in\ social\ network\}$
$H_1 = \{There\ is\ peak\ points\ in\ social\ network\}$

$$D(I, I_n) = \sum |I_n - \bar{I}| \tag{2}$$

First, $D$ ranks informed nodes based on their differences (Eq. (2)). It decides how much a new instance is different from the others. When instance is farther from the mean of the number of informed nodes, $D$ is high. In the next step, a statistic is specified to rank $D$ of new instance with respect to the others. This statistic is defined in Eq. (3):

$$p-value = \frac{\#\{i : D_i > D_n\} + \theta_n \#\{i : D_i = D_n\}}{n} \tag{3}$$

The changes of $p\_value$ into higher values can be considered as number of informed nodes are running away from their representative, where $\theta_n$ is picked from [0, 1] randomly. In contrast, having data close to their mean bring across the meaning that $p\_values$ are approaching smaller values. In order to decide whether $H_0$ must be accepted or not, a martingale in Eq. (4) is defined according to the sequence of $p\_values$.

$$M_i^{(\varepsilon)} = \varepsilon p_i^{\varepsilon-1} M_{i-1}^{(\varepsilon)} \qquad (4)$$

According to Doob's Maximal Inequality [8], it is improbable for $M_k$ to have a high value. So, we can find changes when martingale value is greater than $\lambda$ [27].

## 4. Experimental Results

In order to precisely cover all the important results and observations, this section is broken down into two subsections. First, we describe our experimental datasets and second, we describe results of running martingale approach on these datasets.

### 4.1 Datasets

In order to evaluate the performance of our proposed algorithm to detect peak points of news in social networks, we use two popular datasets in modeling patterns of information diffusion.

The first one is Memetracker dataset [11]. The Memetracker dataset uses a methodology to select the most popular phrases within blog posts and news articles. This dataset contains 17 million different phrases which appeared in blog posts and news articles. By using the mentioned methodology [11] 343 million textual phrases were chosen from more than 1 million online social media from September 1, 2008 to August 31, 2009[1]. We have used 1000 short textual phrases from Memetracker dataset with the biggest volume in 5- days' window around their peak.

Our second dataset is Twitter hashtags. Twitter is one of the most popular social networks around the globe. The selected posts were chosen from June 1, 2009 to December 31, 2009[2]. During that time 6 million Twitter hashtags detected and 1000 of them were chosen with the biggest volume during 5-days window around their peak.

Table 1: List of datasets used in this study and their size and duration

| Datasets | Time Duration | Number |
|---|---|---|
| Memetracker | Sep1, 2008 – Aug31, 2009 | 1000 Quotes and Phrases |
| Twitter | Jun1, 2009 – Dec31, 2009 | 1000 Twitter Hashtags |

### 4.2 Results and Analysis

We ran our proposed algorithm on mentioned datasets and following results are obtained. We compare the prediction peak points by martingale with the real peak points of each news in social networks. We plot the patterns of information diffusion of each news to show the peak points. Due to the space limitation in this paper we show only the results on 3 major phrase of Memetracker and 5 important hashtags of Twitter. Table 2 lists the used phrases of Memetracker and table 3 shows the list of Twitter hashtags.

Figures 1 to 3 show the patterns of information diffusion for phrases of Memetracker dataset. Running the proposed algorithm on each news can predict the peak points of that news.

Table 2: List of phrase of each plot on Memetracker

| Figure | Phrase |
|---|---|
| Fig.1 | "They don't care about us" |
| Fig.2 | "The most serious financial crisis" |
| Fig.3 | "I cannot make you love me" |

Figure 1 shows the patterns of information diffusion on phrase "they don't care about us". We can see the peak points of this news are around time 40 and our algorithm predict this time properly.

---

[1] http://snap.stanford.edu/data/volumeseries.html
[2] http://snap.stanford.edu/data/volumeseries.html

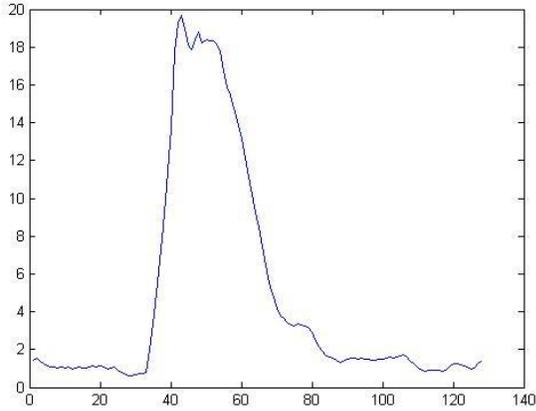

Fig. 1 Result of running proposed algorithm on "They don't care about us". Obviously real first peak points is at time 40 and first alert of our method is at time 40 too. Martingale method can predict peak points with little difference of real one. Second alert is at time 70, as the plot represent there is small peak point near time 70. This method has some error to detect another peak points around time 50.

Phrase "the most serious financial crisis" have two peak points around the time 40. Figure 2 shows the pattern of this phrase and the proposed algorithm alerts at time 47. Based on this pattern the proper prediction is obvious.

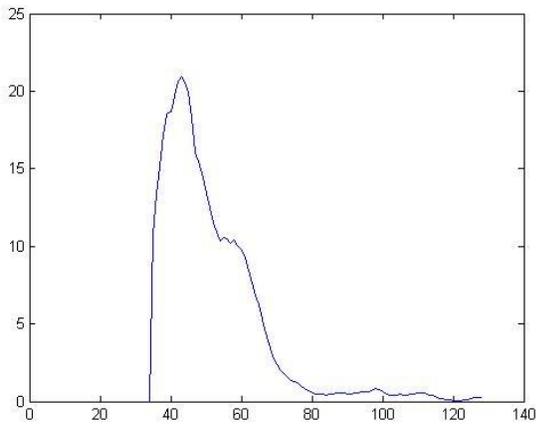

Fig. 2 Result of running proposed algorithm on "The most serious financial crisis". Our approach detected peak point at time 47. This plot shows correct detection for this news. Time 47 is the time when this news reached the peak of its popularity.

In figure 3, the pattern of diffusion for phrase "I cannot make you love me" shows the peak points. Our martingale approach alerts at times 25 and 50. This prediction is same as the real peak point of this news.

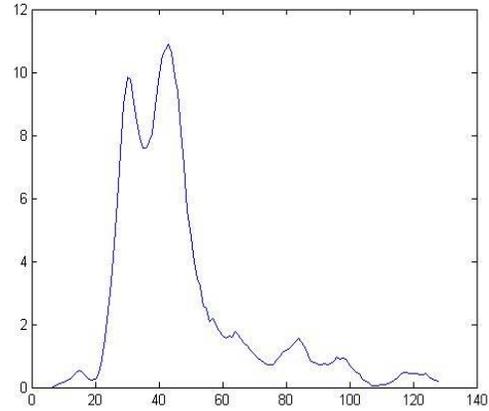

Fig. 3 Result of running proposed algorithm on "I cannot make you love me". This pattern has two big peak points. First one is at time 25 and second one is in around time 50. Martingale approach alerts at two times 25 and 50. One of the most important features of martingale approach is its online detection.

We also use our approach on Twitter dataset to show the effectiveness of proposed algorithm. Figure 4 to 9 illustrate the results of running martingale approach to detect the peak points of hashtags of Twitter.

Table 3: List of hashtags of each plot on Twitter

| *Figure* | *Hashtags* |
|---|---|
| Fig.4 | #iranelection |
| Fig.5 | #bestadvice |
| Fig.6 | #iadmit |
| Fig.7 | #imtiredof |
| Fig.8 | #truth |
| Fig.9 | #turnon |

Figure 4 shows the result of our approach on Twitter hashtag "iranelection". Martingale approach alerts at times 25, 40 and 111. The real peak points are obvious in figure 4.

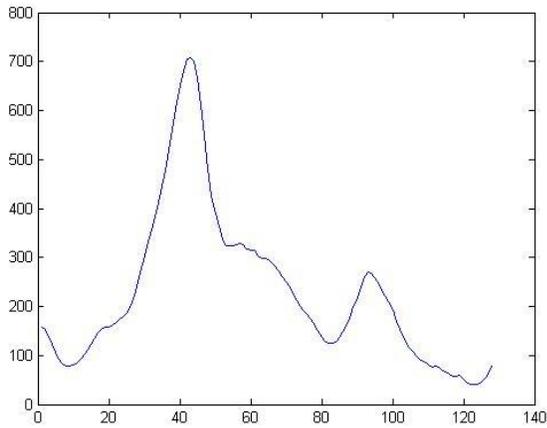

Fig. 4 Result of running proposed algorithm on "#iranelection". The highest peak point of this news is around time 40 and martingale approach predict this time correctly. Our algorithm alerts in times 25 and 111 too. Around time 20 a short peak point can be seen. But time 111 have some differences with the real peak point in time 95. This approach can detect peak points with a little difference toward real peak points online.

Figure 5 illustrates the peak point's time prediction of "#bestadvice" hashtag. The real peak points for this news are around times 40 and 70, but proposed algorithm predict these times 40 and 103.

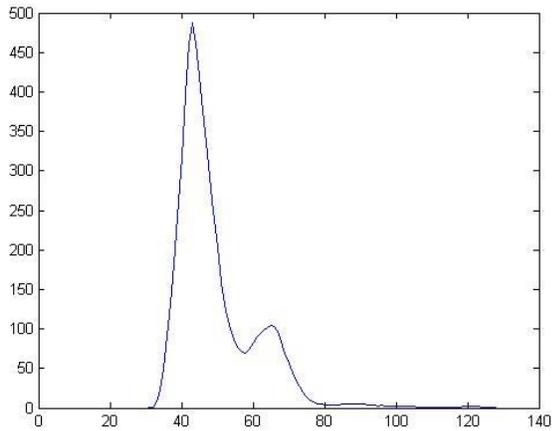

Fig. 5 Result of running proposed algorithm on "#bestadvice". This news has two peak points. First is around time 40 and second is time 70. Martingale method alerts at times 40 and 103. First detection is true but the second is erroneous.

Peak point for "#iadmit" hashtag predicts at time 40. This detection is true as we can see in figure 6.

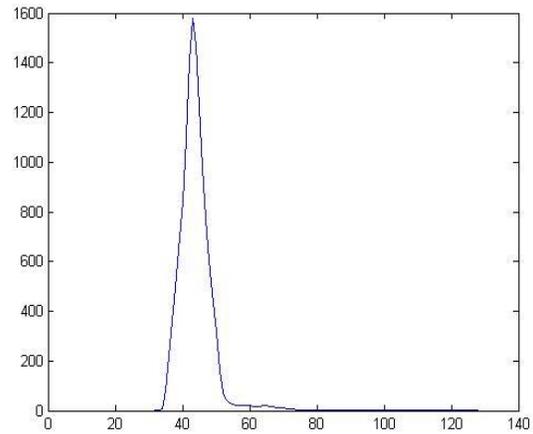

Fig. 6 Result of running proposed algorithm on "#iadmit". Martingale approach alerts at time 40. As it is obvious from the above diffusion pattern real peak points in time 40. Proposed method detect this time correctly.

The results of running proposed method on "#imtiredof" hashtag is shown in figure 7. The real peak points are at around times 40 and 70 and our algorithm predicts times 40 and 77.

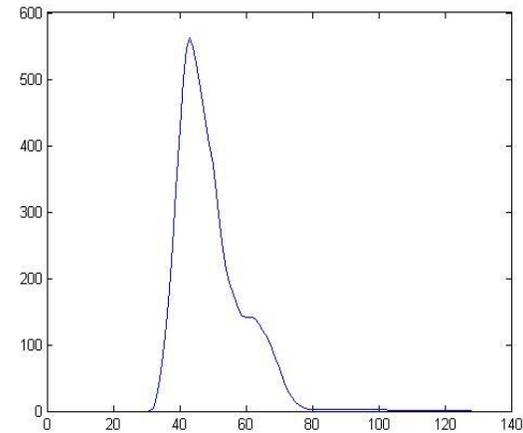

Fig. 7 Result of running proposed algorithm on "#imtiredof". Online predictions of martingale approach are times 40 and 77. Detected times are the true times of peak points for this news.

The prediction peak points for "#truth" hashtag are times 25 and 44. Figure shows the information diffusion patterns of this news. Our system predicts these times correctly.

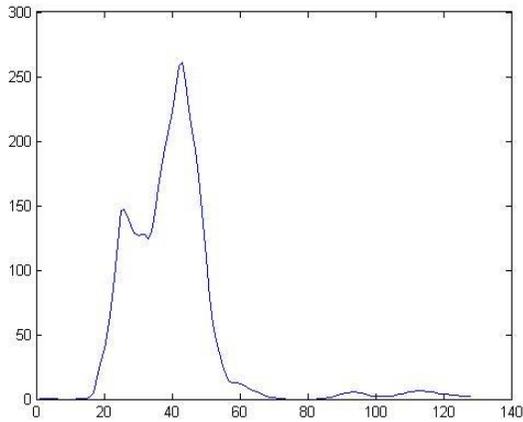

Fig. 8 Result of running proposed algorithm on "#truth". Peak points are occurred at around times 20 and 40. Our approach alerts in time 25 and 40. Seems it is a true detection. Our method work properly to predict these peak points.

Figure 9 shows the real peak points of "#turnon" hashtag. Martingale approach detects 41 and 80 as peak points. But this hashtag has only one peak point.

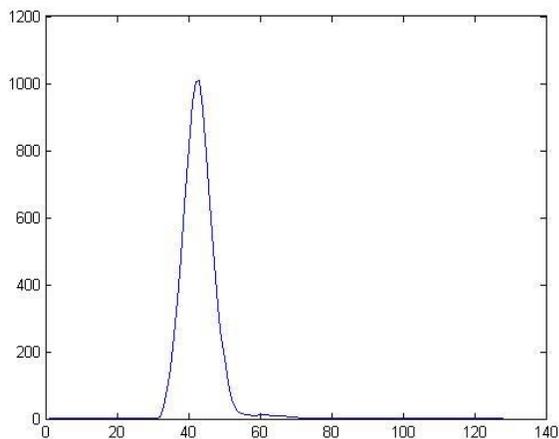

Fig. 9 Result of running proposed algorithm on "#turnon". This news has one peak points at time 41. Martingale approach predicts peak points for this hashtag with error because of alerting at times 41 and 80. The second detection is our approach fault.

News like hashtags or memes spread over social networks and some of them have early start but decays over the times and the others spread slowly. In fact, there are different diffusion patterns for news. But all of them have peak points when they reach the peak of their popularity over time.

From the all above results we can understand our algorithm detects the peak points of news correctly, however this method has some noise in prediction. Solving this problem is our next task for future.

## 5. Conclusion

Understanding social networks structure and how information propagates through them is one of the challenging tasks for researchers of this scope. Studying patterns of information diffusion gives us some useful knowledge on behavior of news in social networks. In this paper, we have proposed a new approach to online detected peak points of news spreading in these networks. Our Algorithm use martingale approach for this prediction. A sequence of infected nodes is given to this algorithm as inputs. Experimental results and analysis show the ability of our proposed martingale approach to online predict the peak points properly.

**S.Babakhani** is the MSc student in information technology (management information system) of CSE and IT Department of Shiraz University. As one of her research interests, she focuses on social networks specially information diffusion domain.

**N.Mozaffari** is the Ph.D. student in artificial intelligence of CSE and IT Department of Shiraz University. As one of her research interests, she focuses on social networks specially information diffusion domain.

**A. Hamzeh** received his Ph.D. in artificial intelligence from Iran University of Science and Technology (IUST) in 2007. Since then, he has been working as assistant professor in CSE and IT Department of Shiraz University. There, he is one of the founders of local CERT center which serves as the security and protection service provider in its local area. As one of his research interests, he recently focuses on cryptography and steganography area and works as a team leader in CERT center to develop and break steganography method, especially in image spatial domain. Also, he works as one of team leaders of Soft Computing group of Shiraz University working on bio-inspired optimization algorithms. He is co-author of several articles in security and optimization.